\documentclass{llncs}
\usepackage[utf8]{inputenc}
\usepackage{microtype}
\usepackage{pifont}

\usepackage{tikz}
\usetikzlibrary{positioning,shapes,petri}
\usetikzlibrary{automata}
\usetikzlibrary{arrows,backgrounds,positioning,fit,calc}
\usetikzlibrary{matrix}

\usepackage{xspace} 
\usepackage{hyperref}
\usepackage{macros}

\title{Parameterized verification of synchronization in constrained
  reconfigurable broadcast networks\thanks{This work has been supported by
    the Indo-French research unit UMI Relax, and by ERC project EQualIS~(308087).}}
\author{
{Balasubramanian A.R.}\inst{1} \and Nathalie Bertrand\inst{2} \and Nicolas Markey\inst{2}}
\institute{Chennai Mathematical Institute -- Chennai (India)
  \and
  Univ. Rennes, Inria, CNRS, IRISA -- Rennes (France)}

\usepackage{graphicx}
\usepackage{array}
\usepackage{url}
\usepackage{xcolor}
\usepackage{amsmath, amsfonts, amssymb,todonotes}
\usepackage{enumitem}
\usepackage{ifthen}

\usepackage{thm-restate}

\newif\ifCone
\newif\ifCtwo
\newif\ifCthree
\newif\ifCfour
\newif\ifCfive

\begin{document}
\hfuzz=2pt
\def\floatpagefraction{.9}
\def\textfraction{.1}
\def\topfraction{.9}

\maketitle

\begin{abstract}
Reconfigurable broadcast networks provide a convenient formalism for
modelling and reasoning about networks of mobile agents broadcasting
messages to other agents following some (evolving) communication
topology. The~parameterized verification of such models aims at
checking whether a given property holds irrespective of the initial
configuration (number of agents, initial states and initial
communication topology).
We~focus here on the synchronization property, asking whether all
agents converge to a set of target states after some execution. This
problem is known to be decidable in polynomial time when no constraints
are imposed on the evolution of the communication topology (while
it~is undecidable for static broadcast networks).

%

In this paper we investigate how various constraints on
reconfigurations affect the decidability and complexity of the
synchronization problem. In~particular, we~show that when bounding the
number of reconfigured links between two communications steps by a constant,
synchronization becomes undecidable; on~the other hand,
 synchronization remains decidable in \PTIME\ when the bound grows with
the number of agents.

\end{abstract}

\section{Introduction}

There are numerous application domains for
networks formed of an arbitrary number of anonymous agents executing
the same code: prominent examples are distributed algorithms,
communication protocols, cache-coherence protocols, and biological
systems such as populations of cells or
individuals,~etc. The~automated verification of such systems is
challenging~\cite{Suz-ipl88,GS92,Esp14,Bloem-book}: its~aim is to validate at once all
instances of the model, independently of the (parameterized) number of
agents.
Such a problem can be phrased in terms of infinite-state-system
verification. Exploiting symmetries may lead to efficient algorithms
for the verification of relevant properties~\cite{ES96}.


Different means of interactions between agents can be considered in
such networks, depending on the application domain. Typical
examples are shared variables~\cite{Hag11,EGM13,BMRSS16},
\emph{rendez-vous}~\cite{GS92},
and broadcast communications~\cite{EFM99,DSZ10}.
In this paper, we target ad~hoc networks~\cite{DSZ10},
in which
the agents can broadcast  messages simultaneously to all their
neighbours, \emph{i.e.}, to all the agents that are within their radio
range. The~number of agents and the communication topology are fixed
once and for all at the beginning of the execution. Parameterized
verification of broadcast networks checks if a specification is met
independently of the number of agents and communication topology.
It~is usually simpler to reason about 
the dual problem of the existence of an initial
configuration (consisting of a network size, an initial state for each
agent, and a communication topology) from which some execution
violates the given specification.

Several types of specifications have been considered 
in the literature. We~focus here on coverability and synchronization:
\emph{does there exist an initial
  configuration from which
  some agent (resp.~all agents at the same time) may reach a
  particular set of target states}.
Both problems are undecidable; decidability of coverability can be
regained by bounding the length of simple paths in the communication
topology~\cite{DSZ10}.

In the case of mobile ad~hoc networks~(MANETs), agents are mobile, so
that the communication links (and thus the neighbourhood of each
agent) may evolve over time. To~reflect the mobility of agents,
Delzanno~\emph{et~al.} studied \emph{reconfigurable} broadcast
networks~\cite{DSZ10,DSTZ12}. In~such networks, the~communication
topology can change arbitrarily at any time.
Perhaps surprisingly, this modification not only allows
for a more faithful modelling of MANETs, but it also leads to
decidability of both the coverability and the synchronization
problems~\cite{DSZ10}. A~probabilistic extension of reconfigurable
broadcast networks has been studied in~\cite{BFS14,BFS15} to model
randomized protocols.

\looseness=-1
A drawback of the semantics of reconfigurable broadcast networks is
that they allow arbitrary changes at each reconfiguration.
Such arbitrary reconfigurations may not be realistic, especially in settings
where communications are frequent enough, and mobility is slow
and not chaotic.
In~this paper, we~limit the
impact of reconfigurations in several ways, and study how those
limitations affect the
decidability and complexity of
parameterized verification of synchronization.



More specifically, we~restrict reconfigurations by limiting the number
of changes in the communication graph, either by considering
\emph{global} constraints (on the total number of edges being
modified), or by considering \emph{local} constraints (on the number
of updates affecting each individual node). We~prove that
synchronization is decidable when imposing constant local constraints,
as well as when imposing global constraints depending (as a divergent
function) on the number of agents. On~the other hand, imposing a
constant global bound makes synchronization undecidable. We~recover
decidability by bounding the maximal degree of each node by~$1$.


\section{Broadcast networks with constrained reconfiguration}
In this section, we~first define reconfigurable broadcast networks;
we~then introduce several constraints on reconfigurations along
executions, and investigate how they compare one to another and with
unconstrained reconfigurations.

\subsection{Reconfigurable broadcast networks}
\begin{definition}
  A \emph{broadcast protocol} is a tuple
  $\BP =(\States,\initStates,\Mess,\Trans)$ where $\States$ is a finite set of control
  states; $\initStates \in \States$ is the set of initial control
  states;
  $\Mess$~is a finite alphabet; and
  $\Trans \subseteq (\States \times
  \set{\broadcast{\mess},\receive{\mess} \mid \mess \in \Mess }\times
  \States)$ is the transition relation.
\end{definition}


A~(reconfigurable) broadcast network is a system made of several copies
of a single broadcast protocol~$\BP$.
Configurations of such a network are
undirected graphs whose each node is labelled with a state
of~$\BP$. Transitions between configurations can either be
reconfigurations of the communication topology (\emph{i.e.}, changes
in the edges of the graph), or a communication via broadcast of a
message (\emph{i.e.}, changes in the labelling of the graph).
Figures~\ref{fig:ex-1constrVSmobile} and~\ref{fig-exexec} respectively
display an example of a broadcast protocol and of an execution of a
network made of three copies of that protocol.

\begin{figure}[btp]
\begin{center}
\begin{tikzpicture}[shorten >=1pt,node distance=7mm and 2cm,on grid,auto,semithick]
  \node[state,inner sep=1pt,minimum size=5mm] (q_0) {$\state_0$};
  \path (q_0.-135) edge[latex'-] ++(-135:4mm);
\node[state,inner sep=1pt,minimum size=5mm] (q_1) [right = of q_0] {$\state_1$};
\node[state,inner sep=1pt,minimum size=5mm] (q_2) [right = of q_1] {$\state_2$};
\node[state,inner sep=1pt,minimum size=5mm] (q_3) [right = of q_2] {$\state_3$};
\node[state,double,inner sep=1pt,minimum size=5mm] (q_4) [right = of q_3] {$\state_4$};
\path[-latex']
(q_0) edge node {$\broadcast{\amess}$} (q_1)
(q_1) edge node {$\broadcast{\bmess}$} (q_2)
(q_2) edge node {$\receive{\cmess}$} (q_3)
(q_3) edge node {$\receive{\dmess}$} (q_4);

\node[state,inner sep=1pt,minimum size=5mm] (q_5) [above = of q_1] {$\state_5$};
\node[state,double,inner sep=1pt,minimum size=5mm] (q_6) [right = of q_5] {$\state_6$};
\path[-latex']
(q_0) edge[bend left] node[above] {$\receive{\amess}$} (q_5)
(q_5) edge node {$\broadcast{\cmess}$} (q_6);
\node[state,inner sep=1pt,minimum size=5mm] (q_7) [below = of q_1] {$\state_7$};
\node[state,double,inner sep=1pt,minimum size=5mm] (q_8) [right = of q_7] {$\state_8$};
\path[-latex']
(q_0) edge[bend right] node[below] {$\receive{\amess}$} (q_7)
(q_7) edge node {$\broadcast{\dmess}$} (q_8);
\end{tikzpicture}
\caption{Example of a broadcast protocol}
\label{fig:ex-1constrVSmobile}
\end{center}
\par

\def\step#1#2#3#4#5#6#7#8#9{%
  \draw[dotted,rounded corners=3mm] (0,1.4) -| (.4,-1.4)
  node[coordinate,pos=.75] (front) {}
  node[coordinate,pos=.8] (frontT) {}
  node[coordinate,pos=.65] (frontB) {}
  -| (-.4,1.4)
  node[coordinate,pos=.75] (rear) {} -- (0,1.4);
  \ifCthree\path (frontT) edge[-latex',dotted] node[below] {$#9$} +(0:6mm);\fi
  \draw (0,0) node[lrond5,opacity=.#2,fill=black!80!white] (a) {};
  \draw (a) node[lrond5] {$#1$};
  \draw (0,1) node[lrond5,opacity=.#4,fill=black!80!white] (b) {};
  \draw (b) node[lrond5] {$#3$};
  \draw (0,-1) node[lrond5,opacity=.#6,fill=black!80!white] (c) {};
  \draw (c) node[lrond5] {$#5$};
  \ifCone\path (a) edge[dashed] (b);\fi
  \ifCtwo\path (a) edge[dashed] (c);\fi
}

\Conetrue
\Ctwotrue
\Cthreetrue 
\begin{center}
  \begin{tikzpicture}
    \begin{scope}
      \step{q_0}0{q_0}0{q_0}0{}{}B
    \end{scope}
    \begin{scope}[xshift=1.5cm]
      \step{q_1}1{q_5}1{q_7}1aaR
      \path (a) node[left=2mm] {$\broadcast a$};
      \path (b) node[left=2mm] {$\receive a$};
      \path (c) node[left=2mm] {$\receive a$};
    \end{scope}
    \Conefalse
    \Ctwofalse
    \begin{scope}[xshift=3cm]
      \step{q_1}0{q_5}0{q_7}0{}{}B
    \end{scope}
    \begin{scope}[xshift=4.5cm]
      \step{q_2}1{q_5}0{q_7}0abR
      \path (a) node[left=2mm] {$\broadcast b$};
    \end{scope}
    \Conetrue
    \begin{scope}[xshift=6cm]
      \step{q_2}0{q_5}0{q_7}0{}{}B
    \end{scope}
    \begin{scope}[xshift=7.5cm]
      \step{q_3}1{q_6}1{q_7}0bcR
      \path (b) node[left=2mm] {$\broadcast c$};
      \path (a) node[left=2mm] {$\receive c$};
    \end{scope}
    \Ctwotrue
    \begin{scope}[xshift=9cm]
      \step{q_3}0{q_6}0{q_7}0{}{}B
    \end{scope}
    \Cthreefalse
    \begin{scope}[xshift=10.5cm]
      \step{q_4}1{q_6}0{q_8}1cdR
      \path (c) node[left=2mm] {$\broadcast d$};
      \path (a) node[left=2mm] {$\receive d$};
    \end{scope}
  \end{tikzpicture}
\caption{Sample execution under reconfigurable semantics, synchronizing to $\set{\state_4,\state_6,\state_8}$ ($B$-transitions are communications steps, $R$ are reconfiguration steps).}
\label{fig:ex-exec-mobile}\label{fig-exexec}
\end{center}
\end{figure}

Formally, we first define undirected labelled graphs. Given a
set~$\Labels$ of labels, an \emph{$\Labels$-graph} is an undirected
graph $\graph = (\Nodes,\Edges,\labelf)$ where $\Nodes$ is a finite
set of nodes; $\Edges \subseteq \Part_2(\Nodes)$\footnote{For a finite
  set~$S$ and $1\leq k\leq \size S$, we~let $\Part_k(S)=\{ T\subseteq
  S\mid \size T=k\}$.} (notice in particular that such a graph has no
self-loops);
finally, $\labelf\colon \Nodes \to \Labels$ is the labelling
function. We~let $\Graphs_\Labels$ denote the (infinite) set of
$\Labels$-labelled graphs. Given a graph $\graph \in \Graphs_\Labels$,
we~write $\node \sim \node'$ whenever $\{\node,\node'\} \in \Edges$
and we let $\Neigh{\graph}{\node} =\{\node' \mid \node \sim \node'\}$
be the neighbourhood of~$\node$, \emph{i.e.}\ the set of nodes
adjacent to~$\node$. For a label~$\alabel$, we~denote
by~$\alabelnb{\graph}{\alabel}$ the number of nodes in $\graph$
labelled by~$\alabel$. Finally $\labelf(\graph)$ denotes the set of
labels appearing in nodes of $\graph$.

The semantics of a reconfigurable broadcast network based on broadcast
protocol $\BP$ is an infinite-state transition system
$\TS(\BP)$. The~configurations of $\TS(\BP)$ are $\States$-labelled
graphs. Intuitively, each node of such a graph runs protocol~$\BP$,
and may send\slash receive messages to\slash from its neighbours.
A~configuration $(\Nodes,\Edges,\labelf)$ is said \emph{initial} if
$\labelf(\Nodes)\subseteq \initStates$.
From a configuration $\graph = (\Nodes,\Edges,\labelf)$, two types of
steps are possible. More precisely, there is a step
from~$(\Nodes, \Edges, \labelf)$ to $(\Nodes',\Edges',\labelf')$ if
one of the following two conditions holds:
\begin{description}
\item[(reconfiguration step)] $\Nodes' =\Nodes$ and $\labelf' =\labelf$:
  a~reconfiguration step does not change the
  set of nodes and their labels, but may change the edges arbitrarily;
\item[(communication step)] $\Nodes'=\Nodes$, $\Edges'= \Edges$, and there
  exists $\node \in \Nodes$ and $\mess \in \Mess$ such that 
  $(\labelf(\node),\broadcast{\mess},\labelf'(\node)) \in \Trans$, and
  for every $\node'$, if $\node' \in \Neigh{\graph}{\node}$, then
  $(\labelf(\node'),\receive{\mess},\labelf'(\node')) \in \Trans$,
  otherwise $\labelf'(\node') = \labelf(\node')$: a~communication step
  reflects how nodes evolve when one of them broadcasts a message to
  its neighbours.
\end{description}
An \emph{execution} of the reconfigurable broadcast network is a
sequence $\exec=(\graph_i)_{0\leq i\leq r}$ of configurations such
that for any~$i<r$, there is a step from~$\graph_i$ to~$\graph_{i+1}$
and $\exec$ strictly alternates communication and reconfiguration
steps (the latter possibly being trivial).
An~execution is \emph{initial} if it starts from an initial
configuration.

An important ingredient that we heavily use in the sequel
is \emph{juxtaposition} of configurations
and \emph{shuffling} of executions.  The juxtaposition of two
configurations $\graph = (\Nodes,\Edges,\labelf)$ and
$\graph' = (\Nodes',\Edges',\labelf')$ is the configuration
$\graph \oplus \graph' = (\Nodes \uplus \Nodes',\Edges \uplus
\Edges',\labelf_\oplus)$, in which $\labelf_\oplus$ extends both
$\labelf$ and~$\labelf'$: $\labelf_\oplus(\node) = \labelf(\node)$ if
$\node \in \Nodes$ and $\labelf_\oplus(\node) = \labelf'(\node)$ if
$\node \in \Nodes'$. We~write $\graph^2$ for the juxtaposition
of~$\graph$ with itself, and, inductively, $\graph^N$ for the
juxtaposition of $\graph^{N-1}$ with~$\graph$.
A~shuffle of two executions~$\exec=(\graph_i)_{0\leq i\leq r}$ and
$\exec'=(\graph'_j)_{0\leq j\leq r'}$ is an execution $\exec_\oplus$
from~$\graph_0\oplus\graph'_0$ to~$\graph_r\oplus\graph'_{r'}$
obtained by interleaving $\exec$ and~$\exec'$. Note that a
reconfiguration step in $\exec_\oplus$ may be composed of
reconfigurations from both $\exec$ and~$\exec'$.
We~write $\exec \oplus \exec'$ for the set of shuffle
executions obtained from~$\exec$ and~$\exec'$.

Natural decision problems for reconfigurable broadcast networks
include checking whether some node may reach a target state, or
whether all nodes may synchronize to a set of target states. More
precisely, given a broadcast protocol~$\BP$ and a subset
$\targetset \subseteq \States$, the \emph{coverability problem} asks
whether there exists an initial execution~$\exec$ that visits a
configuration~$\graph$ with $\labelf(\graph) \cap F\not=\emptyset$,
and the \emph{synchronization problem} asks whether there exists an
initial execution~$\exec$ that visits a configuration~$\graph$ with
$\labelf(\graph) \subseteq F$.
%
%
For unconstrained reconfigurations, we~have:

\begin{theorem}[\cite{DSZ10,DSTZ12,Fou15}]\label{thm-mobile}
  The coverability and synchronization problems are decidable in \PTIME for
  reconfigurable broadcast protocols.
\end{theorem}

\begin{remark}
  The synchronization problem was proven decidable
  in~\cite{DSZ10}, and \PTIME membership was given
  in~\cite[p.~41]{Fou15}. The~algorithm consists in computing the set
  of states of~$\BP$ that are both reachable (\emph{i.e.}, coverable)
  from an initial configuration and co-reachable from a target
  configuration.  This can be performed by applying iteratively the
  algorithm of~\cite{DSTZ12} for computing the set of reachable states
  (with reversed transitions for computing co-reachable states).
%
\end{remark}

\begin{example}\label{example-3branches}
  Consider the broadcast protocol of
  Fig.~\ref{fig:ex-1constrVSmobile} with $\initStates = \{\state_0\}$.
  From each state, unspecified message receptions lead to an (omitted)
  sink state; this way, each broadcast message triggers a transition
  in all the neighbouring copies.

  \looseness=-1
For that broadcast protocol, one easily sees that it is possible to
synchronize to the set $\set{\state_4,\state_6,\state_8}$. Moreover,
three copies are needed and sufficient for that objective, as
witnessed by the execution of Fig.~\ref{fig:ex-exec-mobile}. The
initial configuration has three copies and two edges. If the central
node broadcasts~$\amess$, the other two nodes receive, one proceeding
to~$\state_5$ and the other to~$\state_7$.  Then, we assume the
communication topology is emptied before the same node
broadcasts~$\bmess$, moving to~$\state_2$. Finally the node
in~$\state_5$ connects to the one in~$\state_2$ to communicate
on~$\cmess$ and then disconnects, followed by a similar communication
on~$\dmess$ initiated by the node in~$\state_7$.
\end{example}

\subsection{Natural constraints for reconfiguration}
Allowing arbitrary changes in the network topology may look unrealistic.
In~order to address this issue, we introduce several ways of bounding the
number of reconfigurations after each communication step.
For~this,
we~consider the following natural pseudometric between graphs, which
for simplicity we call \emph{distance}.

\begin{definition}
  Let $\graph = (\Nodes,\Edges,\labelf)$ and
  $\graph' = (\Nodes',\Edges',\labelf')$ be two $\Labels$-labelled
  graphs. The distance between $\graph$ and $\graph'$ is defined
  as
\[
 \dist{\graph}{\graph'} = 
 \lsize{E \cup E' \setminus (E \cap E')}
\]
  when $\Nodes = \Nodes'$ and $\labelf = \labelf'$, and
  $\dist{\graph}{\graph'} = 0$ otherwise.
\end{definition}
Setting the ``distance'' to $0$ for two graphs that do not agree on
the set of nodes or on the labelling function might seem strange at
first. This choice is motivated by the definition of constraints on
executions (see below) and of the number of reconfigurations along an
execution (see Section~\ref{subsec:charac-kconstr}).
Other distances may be of interest in this context; in particular, for
a fixed node~$\node\in\Nodes$, we~let $\ndist{\node}{\graph}{\graph'}$ be the
number of edges involving node~$\node$ in the symmetric difference of
$\Edges$ and~$\Edges'$ (still assuming $\Nodes = \Nodes'$ and $\labelf
= \labelf'$).

\paragraph{Constant number of reconfigurations per step.}
A first natural constraint on reconfiguration consists in bounding
the number of changes in a reconfiguration step by a constant number. Recall that along executions, communication and reconfiguration steps strictly alternate.

\begin{definition}\label{def-kconstr}
  Let $k \in \nats$. An execution $\exec = (\graph_i)_{0\leq i\leq r}$
  of a reconfigurable broadcast network is \emph{$k$-constrained} if
  for every index~$i<r$, it~holds
  ${\dist{\graph_i}{\graph_{i+1}} \leq k}$.
\end{definition}

\addtocounter{example}{-1}
\begin{example}[Contd]
  For the synchronization problem, bounding the number of
  reconfigurations makes a difference. The~sample execution from
  Fig.~\ref{fig:ex-exec-mobile} is not $1$-constrained, and actually
  no $1$-constrained executions of that broadcast protocol can
  synchronize to~$\set{\state_4,\state_5,\state_6}$. This can be shown
  by exhibiting and proving an invariant on the reachable
  configurations (see Lemma~\ref{lemma-1constr}).
\end{example}

\paragraph{Beyond constant number of reconfigurations per step.}
Bounding the number of reconfigurations per step by a constant is
somewhat restrictive, especially when this constant does not depend on
the size of the network.
We~introduce other kinds of constraints here, 
for instance by bounding the number of reconfigurations by~$k$ \emph{on
  average} along the execution, or by having a bound that depends
on the number of nodes executing the protocol.

For a finite execution $\exec = (\graph_i)_{0\leq i\leq r}$ of a
reconfigurable broadcast network, we~write $\nbcomm{\exec}$ for the
number of communication steps along~$\exec$ (notice that $\floor{r/2}
\leq \nbcomm{\exec} \leq \ceil{r/2}$ since we require strict
alternation between reconfiguration and communication steps), and
$\nbreconfig{\exec}$ for the total number of edge reconfigurations
in~$\exec$, that is $\nbreconfig{\exec} = \sum_{i=0}^{r-2}
\dist{\graph_i}{\graph_{i+1}}$.

\begin{definition}
  Let $k\in\nats$.
  An~execution~$\exec$ of a reconfigurable broadcast network
  is said \emph{$k$-balanced} if it starts and ends
with a communication step, and satisfies $\nbreconfig{\exec} \leq k
\cdot (\nbcomm{\exec}-1)$.
\end{definition}
This indeed captures our intuition that along a $k$-balanced
execution, reconfigurations \emph{on average} update less than~$k$
links. 

\medskip
Finally, we~will also consider two relevant ways to constrain
reconfigurations depending on the size of the network: first locally,
bounding the number of reconfigurations \emph{per node} by a constant;
second globally, bounding the total number of reconfigurations by a
function of the number of nodes.

We first bound reconfigurations locally.
\begin{definition}
  Let $k \in \nats$.  An execution $\exec = (\graph_i)_{0\leq i\leq r}$
  of a reconfigurable broadcast network is
  \emph{$k$-locally-constrained}, if, for every node $\node$ and for
  every index $i<r$, $\ndist{\node}{\graph_i}{\graph_{i+1}} \leq k$.
\end{definition}

One may also bound the number of reconfigurations globally using
bounding functions, that depend on the number of nodes in the network:
\begin{definition}
  Let $f \colon \nats \to \nats$ be a 
  function.
An~execution
  $\exec = (\graph_i)_{0\leq i\leq r}$ of a reconfigurable broadcast
  network is \emph{$f$-constrained}, if, writing $n$ for the number of
  nodes in~$\graph_0$, it~holds
  $\dist{\graph_i}{\graph_{i+1}} \leq f(n)$ for any~$i<r$.
\end{definition}
Notice that if $f$ is the constant function $n\in\nats \mapsto k$ for
some~$k\in\nats$, $f$-constrained executions coincide with
$k$-constrained ones, so that our terminology is non-ambiguous.  Other
natural bounding functions are non-decreasing and
\emph{diverging}. This way, the number of possible reconfigurations
tends to infinity when the network size grows, \emph{i.e.}\ $\forall
n.\ \exists k.\ f(k)\geq n$.

\begin{remark}
  Coverability under constrained reconfigurations is easily observed
  to be equivalent to coverability with unconstrained
  reconfigurations: from an unconstrained execution, we~can simply
  juxtapose extra copies of the protocol, which would perform extra
  communication steps so~as~to satisfy the constraint. When dealing
  with synchronization, this technique does not work since the extra
  copies would also have to synchronize to a target state.
  As a consequence, we~only focus on synchronization in the rest of
  this paper.
\end{remark}

\subsection{Classification of constraints}
\looseness=-1
In this section,
we~compare our restrictions.  We~prove that, 
for the synchronization problem, $k$-locally-constrained and
$f$-constrained reconfigurations, for diverging functions~$f$, are
equivalent to unconstrained reconfigurations. On~the other~hand,
we~prove that $k$-constrained reconfigurations are equivalent to
$k$-balanced reconfigurations, and do not coincide with unconstrained
reconfigurations.


\paragraph{Equivalence between unconstrained and
  locally-constrained reconfigurations.}
In~this section, we~handle
the case of $f$-constraints and $k$-local constraints, showing that
they never prevent synchronization.

\begin{lemma}\label{lemma-fconst}
Let $\BP$ be a broadcast protocol, $\targetset \subseteq \States$ be a
target set, and $f$ be a non-decreasing
diverging function. 
If the reconfigurable broadcast network defined by~$\BP$ has an
initial execution synchronizing in~$\targetset$, then it~has an
$f$-constrained initial execution synchronizing in~$\targetset$.
\end{lemma}

\begin{proof}
  We first prove the lemma for the identity function $\Id$. More precisely,
  we prove that for an execution $\exec = (\graph_i)_{0\leq i\leq n}$,
  of the reconfigurable broadcast network, there exists a
  $\Id$-constrained execution $\exec' = (\graph'_j)_{0\leq j\leq m}$,
  whose last transition (if~any) is a communication step, and
  such that for any control state~$q$,
  $\alabelnb{\graph_n}{q} =\alabelnb{\graph_m'}{q}$.
  We~reason by induction on the length of the execution. The claim is
  obvious for ${n = 0}$.  Suppose the property is true for all
  naturals less than or equal to some~$n \in \nats$,
  and consider an execution $\exec=(\graph_i)_{0\leq i\leq n+1}$.
  The~induction hypothesis ensures that there is an $f$-constrained
  execution~$\exec'=(\graph'_j)_{0\leq j\leq m}$ with
  $\alabelnb{\graph_n}{q} = \alabelnb{\graph_m'}{q}$ for all~$q$.  If
  the last transition from $\graph_n$ to~$\graph_{n+1}$ in~$\exec$ is
  a reconfiguration step, then the execution~$\exec'$ witnesses our
  claim.  Otherwise, the transition from $\graph_n$ to~$\graph_{n+1}$
  is a communication step, involving a broadcasting node~$\node$
  of~$\graph_n$ labelled with $\state$, and receiving nodes $\node_1$
  to~$\node_r$ of~$\graph_n$, respectively labelled with $\state_1$
  to~$\state_r$. By~hypothesis, $\graph'_m$ also contains a
  node~$\node'$ labelled with~$\state$ and~$r$ nodes $\node'_1$
  to~$\node'_r$, labelled with~$\state_1$ to~$\state_r$. We~then add
  two steps after~$\graph'_m$ in~$\exec'$: we~first reconfigure the
  graph so that
  $\Neigh{\graph'_{m+1}}{\node'}=\{\node'_i \mid 0\leq i\leq r\}$, which
  requires changing at most $\size{\graph_0}-1$ links, and then
  perform the same broadcast\slash receive transitions as between
  $\graph_n$ and~$\graph_{n+1}$.

  For the general case of the lemma, suppose $f$ is a non-decreasing
  diverging function. Further, let
  $\exec = (\graph_i)_{0\leq i\leq n}$ be an $\Id$-constrained
  execution, and pick~$k$ such that
  $f(k\cdot\size{\graph_0}) \ge \size{\graph_0}$.  Consider the
  initial configuration $\graph_0^k$, made of $k$ copies
  of~$\graph_0$, and the execution, denoted~$\exec^k$, made of $k$
  copies of~$\exec$ running independently from each of the $k$ copies
  of~$\graph_0$ in $\graph_0^k$. Each reconfiguration step involves at
  most $\size{\graph_0}$ links, so that $\exec^k$ is $f$-constrained.
\end{proof}

\begin{restatable}{lemma}{lemmaoneloc}
  \label{lemma-1loc}
Let $\BP$ be a broadcast protocol with $\targetset \subseteq \States$ a 
target set. 
If the reconfigurable broadcast network defined by~$\BP$ has an
initial execution synchronizing in~$\targetset$, then it~has a
$1$-locally-constrained 
initial execution synchronizing in~$\targetset$.
\end{restatable}

\begin{proof}
  Let $\exec = (\graph_i)_{0\leq i\leq n}$ be an execution of the
  reconfigurable broadcast network defined by~$\BP$ that starts and
  ends with a communication step (hence $n$ is~odd) and synchronizing to~$\targetset$.
  
  We write $(s_{2i})_{0\leq 2i\leq n-1}$ for the sequence of communication steps
  in~$\exec$, such that $s_{2i}$ is the step from~$\graph_{2i}$ to~$\graph_{2i+1}$.
  Similarly, $(s_{2i+1})_{1\leq 2i+1\leq n-2}$ is the sequence of
  reconfiguration steps. For each of them, we~let
  $r_{2i+1} = \dist{\graph_{2i+1}}{\graph_{2i+2}}$ be the number of
  reconfigured edges, and $(\edge_{2i+1,j})_{1\leq j\leq r_{2i+1}}$ be
  the sequence (in~arbitrary order) of edges of~$\graph_{2i+1}$
  involved in the reconfiguration step~$s_{2i+1}$.
  We~decompose step~$s_{2i+1}$ into $r_{2i+1}$ \emph{atomic}
  reconfigurations steps $(s_{2i+1}^j)_{1\leq j\leq r_{2i+1}}$.

  Let~$r=\max\{r_{2i+1} \mid 1\leq 2i+1\leq n-2\}$ be the maximum
  number of reconfigured edges in a reconfiguration step in
  $\exec$. We~exhibit a $1$-locally-constrained execution starting
  from configuration~$\graph_0^r$ and ending in~$\graph_n^r$, hence
  synchronizing in~$\targetset$. Each of the $r$ components
  of~$\graph_0^r$ will perform the same sequence of steps as
  along~$\exec$, with the reconfiguration steps being split into
  consecutive atomic reconfiguration steps; the extra components will
  be used to insert communication steps between consecutive
  reconfiguration steps.

  More precisely, the execution decomposes in three parts:
  \begin{itemize}
  \item the first part runs as follows: roughly, it~is made of $r$
    pairs of communication\slash reconfiguration steps. At~step~$1\leq
    i\leq r$, the $i$-th component performs transition~$s_0$; then
    each component $1\leq j\leq i$ performs atomic reconfiguration
    step $s_1^{i+1-j}$ (the other components remain idle).

    In~case some of the first $i$ components are in positions to
    perform a new communication transition, they perform this
    communication step right after the communication step of the
    $i$-th component;
    
  \item the second part is similar, but each of them performs an
    atomic reconfiguration during reconfiguration steps; again, extra
    communication steps may have to be inserted for components that
    are in positions to do so;

  \item the third part starts when the first component performs its
    last communication step. This component has reached its final
    configuration~$\graph_n$, and stops there. The other components
    continue as before until they in turn reach their final
    configuration. This way, the whole system ends up in~$\graph_n^r$.
    \qed
  \end{itemize}
\end{proof}

\paragraph{$k$-constrained and $k$-balanced reconfigurations.}
\label{subsec:charac-kconstr}
We prove here that $k$-constrained and $k$-balanced reconfigurations
are equivalent w.r.t. synchronization, and that they are strictly
stronger than our other restrictions. We~begin with the latter:

\begin{restatable}{lemma}{lemmaoneconstr}
  \label{lemma-1constr}
  There exists a broadcast protocol~$\BP$ and a
  set~$\targetset\subseteq\States$ of target states for which
  synchronization is possible from some initial configuration when
  unconstrained reconfigurations are allowed, and impossible, from
  every initial configuration when only $1$-constrained
  reconfigurations are allowed.
\end{restatable}


\begin{proof}
  For the protocol of Example~\ref{example-3branches}, we~already
  observed the synchronization was possible (with three copies of the
  protocol), and Figure~\ref{fig-exexec} displays an example of a
  $2$-constrained initial execution synchronizing in its target set
  $\{\state_4,\state_6,\state_8\}$.

  We~now prove that no initial configurations can lead to
  synchronization when only using $1$-constrained
  reconfigurations. Assuming otherwise, we~pick a synchonizing
  execution; we~write $x$, $y$ and~$z$ for the number of copies of the
  protocol that end~up in~$\state_4$, $\state_6$ and~$\state_8$,
  respectively. Obviously, the total number of communication steps
  is~$2x+y+z$, so that the total number of reconfigurations must be at
  most~$2x+y+z-1$. Now, when a process arrives in~$\state_5$, it~is
  linked to a process in~$\state_1$, and this link has to be removed
  before message~$\bmess$ can be broadcast. The same holds for those
  processes arriving in~$\state_7$, and a similar argument applies for
  processes entering $\state_2$ and~$\state_3$. It~follows that a
  synchronizing execution has to perform at least $2x+y+z$
  reconfigurations (as~reconfigurations are atomic), which yields a
  contradiction.
\end{proof}

As can be observed in the execution of Fig.~\ref{fig-exexec}, our
definition of $k$-constrained executions is \emph{weak}, and allows
reconfigurations involving less than $k$ edges in the communication
graph; it~even allows trivial reconfigurations, hence consecutive
broadcasts. A~strong version of $k$-constrained executions can be
defined, where \emph{exactly} $k$ edges have to be updated during
reconfiguration phases.

One easily proves that both notions are equivalent (for the
synchronization problem): from a (\emph{weakly}) $k$-constrained
synchronizing execution, one~can build a \emph{strongly}
$k$-constrained synchronizing execution by taking several extra copies
of the graph describing the topology, in which we can add, and
then remove, edges in order to always perform exactly $k$ changes in
the topology.
Actually, we~have:

\begin{lemma}\label{lemma-kto1}
  Let $\BP$ be a broadcast protocol, $\targetset \subseteq \States$ be
  a target set, and $k\in\nats$.
  If the reconfigurable broadcast network defined by~$\BP$ has a
  $1$-constrained initial execution synchronizing in~$\targetset$,
  then it~has a $k$-constrained initial execution synchronizing
  in~$\targetset$.
\end{lemma}

\begin{proof}
Consider a $1$-constrained initial execution~$\exec=(\graph_i)_{0\leq
  i< n}$ synchronizing in~$\targetset$. Let~$m$ be an even number
larger than or equal to~$k+2$.  We~build a $k$-constrained initial
execution $\exec'=(\graph'_j)_{0\leq j< mn}$ by mimicking $m$ copies
of~$\exec$ running in parallel. This execution starts from
$\bigoplus_{0\leq j< m} \graph_0$. Consider a pair of consecutive
steps of~$\exec$, made of a communication step followed by a
$1$-constrained reconfiguration step. It~goes from~$\graph_{2i}$
to~$\graph_{2i+2}$, for some~$i$. We~associate with this pair of steps
$m/2$ pairs of steps from~$\bigoplus_{0\leq j< m}\graph_{2i}$
to~$\bigoplus_{0\leq j< m}\graph_{2i+2}$. For~this, we consider two
copies of~$\graph_{2i}$ at a time (so that there remains at least $k$
auxiliary copies): the first of the two selected copies performs the
communication step, followed by the $1$-constrained reconfiguration
step. In~order to make the latter step $k$-constrained, we perform a
set of $k-1$ (arbitrary) reconfigurations in the remaining $k$
auxiliary copies of~$\graph_{2i}$. Then the second of the two selected
copies performs the communication step, followed by a $1$-constrained
reconfiguration step, which is made $k$-constrained by reversing the
$k-1$ arbitrary reconfigurations performed previously. This way, two
copies have moved to~$\graph_{2i+2}$, which the remaining $m-2$ copies
are still in~$\graph_{2i}$ and can in turn apply the same strategy.

The last communication step of~$\exec$ is handled similarly, but it is
not coupled with a reconfiguration step; we~may need up to $k$
auxiliary copies to perform a $k$-constrained reconfiguration step.
\end{proof}

\begin{corollary}\label{coro-weakconstr}
  Let $\BP$ be a broadcast protocol, $\targetset \subseteq \States$ be
  a target set, and $k\in\nats$.
  The reconfigurable broadcast network defined by~$\BP$ has a (weakly)
  $k$-constrained initial execution synchronizing in~$\targetset$
  (where reconfigurations may change at most $k$ links), if, and
  only~if, it~has a strongly $k$-constrained initial execution
  synchronizing in~$\targetset$ (where each reconfiguration step
  modifies exactly $k$ links).
\end{corollary}

We now prove the main result of this section:
\begin{restatable}{theorem}{thmkconstrkbal}
\label{th:charac-kconstr}
Let $\BP$ be a broadcast protocol and $\targetset\subseteq \States$.
There exists a $k$-constrained initial execution synchronizing
in~$\targetset$ if, and only if, there exists a $k$-balanced initial
execution synchronizing in~$\targetset$.
\end{restatable}

\begin{proof}
  The left-to-right implication is simple: if there is a
  $k$-constrained initial execution synchronizing in~$\targetset$,
  w.l.o.g. we can assume that this execution starts and ends with a
  communication step; moreover, each reconfiguration step contains at
  most $k$ edge reconfigurations, so that the witness execution is
  $k$-balanced. 

  \smallskip
  Let
  $\exec = (\graph_i)_{0\leq i\leq n}$ be a $k$-balanced execution
  synchronizing in~$\targetset$ and starting and ending with
  communication steps (hence $n$ is~odd). We~define the
  potential~$(p_i)_{0\leq i\leq n}$ of~$\rho$ as the sequence of $n+1$
  integers obtained as follows:
  \begin{itemize}
  \item $p_0=0$;
  \item $p_{2i+1} = p_{2i}+k$ for $i\leq (n-1)/2$ (this corresponds to a communication step);
  \item $p_{2i+2} = p_{2i+1} - \dist{\graph_{2i+1}}{\graph_{2i+2}}$
    for $i\leq (n-1)/2-1$  (reconfiguration step).
  \end{itemize}
  That~$\exec$ is $k$-balanced translates as $p_{n-1} \geq 0$: the
  sequence $(p_i)_{0\leq i\leq n}$ stores the value of $k\cdot
  \nbcomm{\rho_{\leq i}} - \nbreconfig{\rho_{\leq i}}$ for each
  prefix~$\rho_{\leq i}$ of~$\rho$; being $k$-balanced means that
  $p_{n}\geq k$, and since the last step is a communication step, this
  in turn means $p_{n-1}\geq 0$.  On~the other hand, in~order to be
  $k$-constrained, it~is necessary (but not sufficient) to have
  $p_i\geq 0$ for all~$0\leq i\leq n$.

  \medskip


  We~build a $k$-constrained execution by shuffling several copies
  of~$\exec$.  We~actually begin with the case where $k=1$, and then
  extend the proof to any~$k$.
%
  We~first compute how many copies we need. For this, we~split~$\rho$
  into several phases, based on the potential~$(p_i)_{0\leq i\leq n}$
  defined above.  A~phase is a maximal segment of~$\exec_{\leq n-1}$
  (the prefix of~$\exec$ obtained by dropping the last (communication)
  step) along which the sign of the potential is constant (or~zero):
  graphs $\graph_i$ and~$\graph_j$ are in the same phase if, and
  only~if, for all $i\leq l\leq l'\leq j$, it~holds $p_l\cdot p_{l'}\geq
  0$. We~decompose~$\exec$ as the concatenation of phases
  $(\exec_j)_{0\leq j\leq m}$; since $\exec$ is $k$-balanced, $m$~is
  even, and $\exec_0$, $\exec_m$,~and all even-numbered phases are
  \emph{non-negative} phases (\emph{i.e.}, the potential is non-negative
  along those executions), while all odd-numbered executions are
  \emph{non-positive} phases.  Also, all phases end with
  potential~zero, except possibly for~$\exec_m$.
  See~Fig.~\ref{fig-phases} for an example of a decomposition into
  phases.

  \begin{restatable}{lemma}{lemmaphase}
    \label{lemma-phase}

    For any phase $\exec_i = \graph_{b_i} \cdots \graph_{e_i}$ of a
    $1$-balanced execution $\exec = \graph_0 \cdots \graph_n$,
    there exists $\kappa_i \leq (e_i-b_i)/2$
    such that for
    any $N \in \nats$, there exists a $1$-constrained execution from
    $\graph_0^{{\kappa_i}} \oplus \graph_{b_i}^N$ to
    $\graph_1^{{\kappa_i}} \oplus \graph_{e_i}^N$.
  \end{restatable}
  
  \begin{proof}
    We~handle non-negative and non-positive phases separately.
    In~a
    non-negative phase, we~name \emph{repeated reconfiguration step}
    any reconfiguration step that immediately follows another
    (possibly from the previous phase) reconfiguration step (so that
    if there are four consecutive reconfiguration steps, the last
    three are said repeated); similarly, we~name \emph{repeated
      communication step} any communication step that is immediately
    followed (possibly in the next phase) by another communication
    step (hence the first three of fours consecutive communication
    steps are repeated). 
    
    We~first claim that any non-negative phase contains at least as
    many repeated communication steps as it contains repeated
    reconfiguration steps. Indeed, any non-repeated communication step
    in a non-negative phase is necessarily followed by a non-repeated
    reconfiguration step, and conversely, and non-negative phases have
    at least as many communication steps as they have reconfiguration
    steps.

    As~a consequence, we~can number all repeated reconfiguration steps
    from~$1$ (earliest) to~$\kappa_i$ (latest), for some~$\kappa_i$,
    and similarly for repeated communication steps.
    Clearly enough, in a non-negative phase,
    for any~$1\leq j\leq \kappa_i$, the repeated
    communication step numbered~$j$ occurs before the repeated
    reconfiguration step carrying the same number.

    We~now build our $1$-constrained execution from
    $\graph_0^{\kappa_i}
 \oplus \graph_{b_i}^N$ to
    $\graph_1^{\kappa_i}
 \oplus \graph_{e_i}^N$.  We~begin
    with a first part, where only the components starting
    from~$\graph_{b_i}$ move:
    \begin{itemize}
    \item the first copy starting in~$\graph_{b_i}$ follows the
      execution~$\exec_i$ until reaching the repeated reconfiguration
      step number~$1$. That reconfiguration step cannot be performed
      immediately as it follows another reconfiguration step. Notice
      that during this stage, this copy has taken at least one
      repeated communication step, numbered~$1$;
    \item the second copy then follows~$\exec_i$ until reaching its
      first repeated communication step (which must occur before the
      first repeated reconfiguration step). It~takes this
      communication step, then allowing the first copy to perform its
      first repeated reconfiguration step;
    \item this simulation continues, each time having the $l+1$-st
      copy of the system taking its $j$-th repeated communication
      step in order to allow the $l$-th copy to perform its $j$-th
      repeated reconfiguration step. Non-repeated steps can always be
      performed individually by each single copy. Also, the first copy
      may always take repeated communication steps not having a
      corresponding reconfiguration step, as in the first stage of
      this part.
    \end{itemize}
    Notice that the number of copies involved in this process is
    arbitrary.  The~process lasts as long as some copies may
    advance within phase~$\exec_i$.
    Hence, when the process~stops, all copies of the original system
    either have reached the end of~$\exec_i$, or are stopped before a
    repeated reconfiguration step. For the copies in the latter
    situation, we~use the copies starting from~$\graph_0$. It~remains
    to prove that having $\kappa_i$ such copies is enough to make
    all processes reach the end of~$\exec_i$.

    For this, we~first assume that the potential associated
    with~$\exec_i$ ends with value~zero. This must be the case of all
    phases except the last one, which we handle after the general
    case.  We~first notice that in the execution we are currently
    building, any repeated communication step performed by any (but
    the very first) copy that started from~$\graph_{b_i}$ is always
    followed by a repeated reconfiguration step. Similarly,
    non-repeated communication steps of any copy is followed by a
    non-repeated broadcast step of the same copy. As~a consequence,
    the potential associated with the global execution we are
    currently building never exceeds the total number of repeated
    communication steps of performed by the first copy; hence it is
    bounded by~$\kappa_i$, whatever the number~$N$ of copies
    involved. As~a consequence, at most $\kappa_i$ communication steps
    are sufficient in order to advance all copies that started
    from~$\graph_{b_i}$ to the end of~$\exec_i$.

    Finally, the case of the last phase $\exec_m$ (possibly ending with positive
    potential) is easily handled, as it has more communication
    steps than reconfiguration steps.

    The proof for non-positive phases is similar.
  \end{proof}

\begin{figure}[t]
  \centering
  \begin{tikzpicture}[scale=.45]
    \tikzstyle{repcom}=[line width=1mm,blue,opacity=.5,loosely dotted]
    \tikzstyle{reprec}=[line width=1mm,red,opacity=.5,densely dotted]
    \fill[black!5!white] (0,-3) -| (8,3) -| (0,-3);
    \fill[black!10!white] (8,-3) -| (18,3) -| (8,-3);
    \fill[black!5!white] (18,-3) -| (23,3) -| (18,-3);
    \draw[dotted] (8,-3) -- +(0,6);
    \draw[dotted] (18,-3) -- +(0,6);
    \path (4,-1.5) node {nonnegative phase};
    \path (13,1.5) node {nonpositive phase};
    \path (20.5,-1.5) node[text width=2cm,align=center] {\centering nonnegative phase};
    \path[use as bounding box] (0,0);
    \draw[-latex']  (0,-3) -- (0,3) node[above] {potential};
    \draw (0,0) -- (24,0);
    \draw (0,0) node[rond,minimum  size=0pt] (0) {}
      -- ++(1,1) node[rond,minimum  size=0pt] (1) {}
      -- ++(1,1) node[rond,minimum  size=0pt] (2) {}
      -- ++(1,1) node[rond,minimum  size=0pt] (3) {}
      -- ++(1,-1) node[rond,minimum  size=0pt] (4) {}
      -- ++(1,-1) node[rond,minimum  size=0pt] (5) {}
      -- ++(1,1) node[rond,minimum  size=0pt] (6) {}
      -- ++(1,-1) node[rond,minimum  size=0pt] (7) {}
      -- ++(1,-1) node[rond,minimum  size=0pt] (8) {} 
      -- ++(1,-1) node[rond,minimum  size=0pt] (9) {}
      -- ++(1,-1) node[rond,minimum  size=0pt] (10) {}
      -- ++(1,1) node[rond,minimum  size=0pt] (11) {}
      -- ++(1,1) node[rond,minimum  size=0pt] (12) {}
      -- ++(1,-1) node[rond,minimum  size=0pt] (13) {}
      -- ++(1,-1) node[rond,minimum  size=0pt] (14) {}
      -- ++(1,1) node[rond,minimum  size=0pt] (15) {}
      -- ++(1,1) node[rond,minimum  size=0pt] (16) {}
      -- ++(1,-1) node[rond,minimum  size=0pt] (17) {}
      -- ++(1,1) node[rond,minimum  size=0pt] (18) {} 
      -- ++(1,1) node[rond,minimum  size=0pt] (19) {}
      -- ++(1,-1) node[rond,minimum  size=0pt] (20) {}
      -- ++(1,1) node[rond,minimum  size=0pt] (21) {}
      -- ++(1,1) node[rond,minimum  size=0pt] (22) {}
      -- ++(1,-1) node[rond,minimum  size=0pt] (23) {}
      -- ++(1,1) node[rond,minimum  size=0pt] (25) {};
      \path (0) edge[style=repcom]
         node[below right=-2mm,black,opacity=1] {\ding{192}} (1) 
      (1) edge[style=repcom]
         node[below right=-2mm,black,opacity=1] {\ding{193}} (2)
      (10) edge[style=repcom]
         node[below right=-2mm,black,opacity=1] {\ding{192}}  (11)
      (14) edge[style=repcom]
         node[below right=-2mm,black,opacity=1] {\ding{193}} (15)
      (17) edge[style=repcom]
         node[below right=-2mm,black,opacity=1] {\ding{194}} (18)
      (20) edge[style=repcom]
         node[below right=-2mm,black,opacity=1] {\ding{192}} (21) ;
    \path (4) edge[style=reprec]
         node[above right=-2mm,black,opacity=1] {\ding{192}} (5)
      (7) edge[style=reprec]
         node[above right=-2mm,black,opacity=1] {\ding{193}} (8)
      (8) edge[style=reprec]
         node[above right=-2mm,black,opacity=1] {\ding{192}} (9)
      (9) edge[style=reprec]
         node[above right=-2mm,black,opacity=1] {\ding{193}} (10)
      (13) edge[style=reprec]
         node[above right=-2mm,black,opacity=1] {\ding{194}} (14)
      ;
  \end{tikzpicture}
  \caption{Phases of a $1$-balanced execution, and correspondence
    between repeated communication steps (loosely dotted blue steps)
    and repeated reconfiguration steps (densely dotted red steps)}
  \label{fig-phases}
\end{figure}
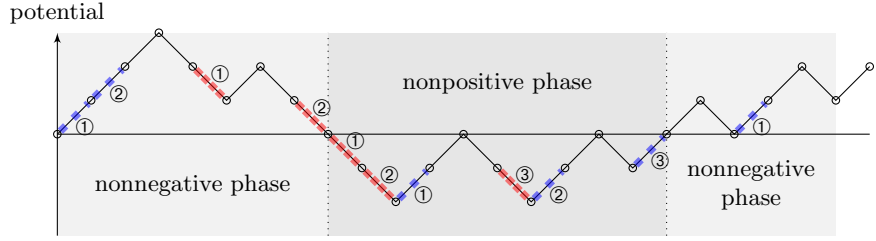

\smallskip
Pick a $1$-balanced execution~$\exec = \graph_0 \cdots \graph_n$, and
decompose it into phases $\exec_1 \cdots \exec_m$. For each
phase~$\exec_i$, we~write $\kappa_i$ for the total number of repeated
reconfiguration steps, and we~let $\kappa=\sum_{1\leq i\leq m}
\kappa_i$ for the total number of repeated reconfiguration steps
along~$\exec$. Notice that $\kappa\leq n/2$.

\begin{restatable}{lemma}{lemmaendphase}
  \label{lemma-endphase}
For every $1$-balanced execution $\exec = \graph_0 \cdots \graph_n$,
and for every $N \in
\nats$, there exists a $1$-constrained execution from
$\graph_1^N \oplus \graph_{e_m}^{\kappa N}$ to
$\graph_n^{N+\kappa N}$.
\end{restatable}

  \begin{proof}
  The copies starting from~$\graph_{e_m}$ only have to perform a
  single communication step in order to reach~$\graph_n$. These steps
  will be used to allow reconfigurations for the $N$ copies
  mimicking~$\exec$. Each of the $N$ copies as to perform at most
  $\kappa$ repeated reconfiguration steps, hence it needs at most
  $\kappa$ external communication steps to reach the end of~$\exec$.
  \end{proof}

Combining the above two lemmas, we obtain the following proposition,
which refines the statement of the Theorem~\ref{th:charac-kconstr}:
\begin{restatable}{proposition}{propphase}
  \label{prop-phase}
  For every $1$-balanced execution $\exec = \graph_0 \cdots \graph_n$
  and every $N \geq \kappa^2+\kappa$, there exists a $1$-constrained
  execution from $\graph_0^N$ to~$\graph_n^N$.
\end{restatable}

  \begin{proof}
  Pick $N\geq \kappa^2+\kappa$, and consider $N$ copies
  of~$\graph_0$. We~take $N-\kappa$ such copies, and apply
  Lemma~\ref{lemma-phase} to each phase~$\exec_i$ of~$\exec$, each
  time involving $\kappa_i$ of the remaining $\kappa$ copies. In~the
  end, we~have $\kappa$ copies in~$\graph_1$ and $N-\kappa$ copies
  in~$\graph_{e_m}$. We~then apply Lemma~\ref{lemma-endphase}, using
  $\kappa^2$ of the copies in~$\graph_{e_m}$ in order to move our
  $\kappa$ copies from~$\graph_1$ to~$\graph_n$. The remaining
  $N-\kappa-\kappa^2$ copies still in~$\graph_{e_m}$ can finally move
  to~$\graph_n$, completing the execution.
\end{proof}

We finally extend this result to $k>1$. In this case,
splitting~$\exec$ into phases is not as convenient as when~$k=1$:
indeed, a non-positive phase might not end with potential~zero (because
communication steps make the potential jump by $k$
units). Lemma~\ref{lemma-phase} would not hold in this case.

We~circumvent this problem by first shuffling $k$ copies of~$\exec$ in
such a way that reconfigurations can be gathered into groups of size
exactly~$k$. This way, we~can indeed split the resulting execution
into non-negative and non-positive phases, always considering
reconfigurations of size exactly~$k$; we~can then apply the techniques
above in order to build a synchronizing $k$-constrained
execution. This completes our proof.\forceqed
\end{proof}

\section{Parameterized synchronization under reconfiguration constraints}
\subsection{Undecidability for $k$-constrained reconfiguration}

Although synchronization is decidable in \PTIME~\cite{DSZ10,Fou15} for
reconfigurable broadcast networks, the problem becomes undecidable
when reconfigurations are $k$-constrained.

\begin{theorem}\label{thm-undec}
  The synchronization problem is undecidable for reconfigurable
  broadcast networks under $k$-constrained reconfigurations.
\end{theorem}

\begin{proof}
  We prove this undecidability result for $1$-constrained
  reconfigurations, by giving a reduction from the halting problem for
  Minsky machines~\cite{Min67}. 
  We~begin with some intuition.
  The state space of our protocol has two types of states:
  \begin{itemize}
  \item \emph{control states} encode the control state of the 2-counter
    machine;
  \item \emph{counter states} are used to model counter values: for
    each counter~$c_j\in \{c_1,c_2\}$, we~have a state~$\zero{j}$ and
    a state~$\one{j}$. The~value of counter~$c_j$ in the simulation
    will be encoded as the number of edges in the communication
    topology between the \emph{control node} and \emph{counter nodes}
    in state~$\one{j}$; moreover, we~will require that control nodes
    have no communication links with counter nodes in state~$\zero
    j$.

    Incrementations and decrementations can then be performed by
    creating a link with a node in~$\zero j$ and sending this node
    to~$\one j$, or sending a $\one j$-node to~$\zero j$ and removing
    the link.
  \end{itemize}
  In~order to implement this, we~have to take care of the facts that
  we may have several control nodes in our network, that we may have
  links between two control nodes or between two counter nodes, or
  that links between control nodes and counter nodes may appear or
  disappear at random.  Intuitively, those problems will be handled as
  follows:
  \begin{itemize}
  \item we cannot avoid having several control nodes; instead, given
    a synchronizing execution of the broadcast protocol, we will
    select one control node and show that it encodes a correct
    execution of the 2-counter machine;
  \item in order to reach a synchronizing configuration, the selected
    control node will have to perform at least as many reconfiguration
    steps as broadcast steps. Because we consider $1$-constrained
    runs, it~will perform exactly the same number of reconfiguration
    steps as broadcast steps, so that no useless\slash unexpected
    reconfigurations may take place during the simulation;
  \item control nodes will periodically run special broadcasts that
    would send any connected nodes (except nodes in state~$\one j$) to
    a sink state, thus preventing synchronization. This way, we~ensure
    that that particular control node is \emph{clean}. Initially,
    we~require that control nodes have no connections at~all.
  \end{itemize}
  \looseness=-1 We~now present the detailed construction, depicted at
  Fig.~\ref{fig-global} to~\ref{fig-aux}. Each state of the protocol
  is actually able to synchronize with all the messages; all nodes
  with no outgoing transitions (i.e., nodes $\zero j'$ and $\done i$,
  which will be part of the target~set) actually carry a self-loop
  synchronizing on all messages; all other transitions that are not
  drawn lead to an (omitted) sink state, which is not part of the
  target set.
  \medskip




    \begin{figure}[ht]
    \centering
    \begin{tikzpicture}[scale=.8]
      \draw (0,0) node[rond] (M0) {$M_0$};
      \draw[latex'-] (M0.-135) -- +(-135:3mm);
      \fill[black!10!white,rounded corners=3mm]
        (2,.6) -| (7.2,-.6) -| (1.4,.6) -- (2,.6);
      \fill[black!10!white,rounded corners=3mm]
        (8,.6) -| (13.2,-1.6) -| (7.4,.6) -- (8,.6);
      \draw (2,0) node[rond] (L1) {$L_1$};
      \draw (6,0) node[rond] (M1) {$M_1$};
      \draw (8,0) node[rond] (L2) {$L_2$};
      \draw (12,0) node[rond] (M2) {$M_2$};
      \draw (12,-1) node[rond] (M'2) {$M'_2$};
      \begin{scope}
      \everymath{\scriptstyle}
      \path (M0) edge[-latex'] node[pos=.35,above] {\broadcast{start}} (L1);
      \foreach \i in {3,4,5} {
        \draw (\i,0) node[dotted,rond3] (N1\i) {};}
      \path (L1) edge[-latex',dotted] (N13);
      \path (N13) edge[-latex',dotted] (N14);
      \path (N14) edge[-latex',dotted] (N15);
      \path (N15) edge[-latex',dotted] (M1);
      \path (M1) edge[-latex'] node[above,pos=.4] {\broadcast{i-exit}} (L2);
      \foreach \i in {9,10,11} {
        \draw (\i,0) node[dotted,rond3] (N2\i) {};
        \draw (\i,-1) node[dotted,rond3] (N'2\i) {};}
      \path (L2) edge[-latex',dotted] (N29);
      \path (L2) edge[-latex',dotted] (N'29);
      \path (N29) edge[-latex',dotted] (N210);
      \path (N'29) edge[-latex',dotted] (N'210);
      \path (N210) edge[-latex',dotted] (N211);
      \path (N'210) edge[-latex',dotted] (N'211);
      \path (N211) edge[-latex',dotted] (M2);
      \path (N'211) edge[-latex',dotted] (M'2);
      \path (M2) edge[-latex',dotted] +(0:1.5cm);
      \path (M'2) edge[-latex',dotted] +(0:1.5cm);
      \end{scope} 
      \path[use as bounding box] (0,0);
      \draw[dotted,shorten <=2mm,shorten >=2mm]
        (2,-.8) edge[out=-160,in=110,looseness=1,-latex'] +(-1.5,-2.3);
      \draw[dotted,shorten <=2mm,shorten >=2mm]
        (13,-1.8) edge[out=-40,in=10,looseness=2,-latex'] +(-1.5,-4.3);
    \end{tikzpicture}
    \caption{Global view of the part of the protocol for control nodes}
    \label{fig-global}

    \bigskip
    \begin{tikzpicture}[scale=.8]
      \begin{scope}[xshift=-.5cm,yshift=-3cm,xscale=.8]
        \fill[black!10!white,rounded corners=3mm] (0,.6) -|
        (17.6,-.6) -| (5,-1.2) -| (-.6,.6) -- (0,.6);
        \path (2.25,-.8) node {\hbox to 3.1cm{\sffamily incrementation module}};
        \draw (0,0) node[rond] (L1) {$L$};
        \draw (2,0) node[rond] (l2) {$\inc1$};
        \draw (4,0) node[rond] (l3) {$\inc2$};
        \draw (6,0) node[rond] (l4) {$\inc3$};
        \draw (8,0) node[rond] (l5) {$\inc4$};
        \draw (10,0) node[rond] (l6) {$\inc5$};
        \draw (12,0) node[rond] (l7) {$\inc6$};
        \draw (14,0) node[rond] (l8) {$\inc7$};
        \draw (16,0) node[rond] (l9) {$M$};
        \begin{scope}
          \everymath{\scriptstyle}
        \path (L1) edge[-latex'] node[above] {\broadcast{i-init}} (l2);
        \path (l2) edge[-latex'] node[above] {\receive{\maux$_1$}} (l3);
        \path (l3) edge[-latex'] node[above] {\broadcast{i-ask$_j$}} (l4);
        \path (l4) edge[-latex'] node[above] {\receive{i-ack$_j$}} (l5);
        \path (l5) edge[-latex'] node[above] {\broadcast{i-ok$_j$}} (l6);
        \path (l6) edge[-latex'] node[above] {\receive{\maux$_2$}} (l7);
        \path (l7) edge[-latex'] node[above] {\receive{\maux$_3$}} (l8);
        \path (l8) edge[-latex'] node[above] {\receive{\maux$_3$}} (l9);
        \path (l9) edge[-latex', shorten >=-3mm] node[above] {\broadcast{i-exit}} +(0:1.6cm);
        \end{scope}
      \end{scope}
      \begin{scope}[xshift=-.5cm,yshift=-5.2cm,xscale=.8]
        \fill[black!10!white,rounded corners=3mm] (0,.6) -| (5.5,1.2) -|
        (13.6,-1.6) -| (-.6,.6) -- (0,.6);
        \path (13,.8) node {\hbox to 0pt{\hss \sffamily decrementation\slash zero-test module}};
        \draw (0,0) node[rond] (L1) {$L$};
        \draw (2,0) node[rond] (l2) {$\dec1$};
        \draw (4,0) node[rond] (l3) {$\dec2$};
        \draw (6,0) node[rond] (l4) {$\dec3$};
        \draw (8,0) node[rond] (l5) {$\dec4$};
        \draw (10,0) node[rond] (l6) {$\dec5$};
        \draw (12,0) node[rond] (l7) {$M$};
        \draw (2,-1) node[rond] (l'2) {$\tst1$};
        \draw (4.5,-1) node[rond] (l'3) {$\tst2$};
        \draw (7,-1) node[rond] (l'4) {$\tst3$};
        \draw (9.5,-1) node[rond] (l'5) {$\tst4$};
        \draw (12,-1) node[rond] (l'6) {$M'$};
        \begin{scope}
          \everymath{\scriptstyle}
          \path (L1) edge[-latex'] node[above] {\broadcast{d-ask$_j$}} (l2);
          \path (l2) edge[-latex'] node[above] {\receive{d-ack$_j$}} (l3);
          \path (l3) edge[-latex'] node[above] {\broadcast{d-ok$_j$}} (l4);
          \path (l4) edge[-latex'] node[above] {\receive{\maux$_4$}} (l5);
          \path (l5) edge[-latex'] node[above] {\receive{\maux$_5$}} (l6);
          \path (l6) edge[-latex'] node[above] {\receive{\maux$_5$}} (l7);
          \path (L1) edge[-latex'] node[below, sloped] {\broadcast{t-ask$_j$}} (l'2);
          \path (l'2) edge[-latex'] node[above] {\receive{t-ack$_j$}} (l'3);
          \path (l'3) edge[-latex'] node[above] {\broadcast{t-ok$_j$}} (l'4);
          \path (l'4) edge[-latex'] node[above] {\receive{\maux$_5$}} (l'5);
          \path (l'5) edge[-latex'] node[above] {\receive{\maux$_5$}} (l'6);
          \path (l7) edge[-latex', shorten >=-3mm] node[above] {\broadcast{d-exit}} +(0:1.6cm);
          \path (l'6) edge[-latex', shorten >=-3mm] node[above] {\broadcast{t-exit}} +(0:1.6cm);
        \end{scope}
      \end{scope}
    \end{tikzpicture}
    \caption{Modules for simulating incrementation and decrementation\slash zero test}
    \label{fig-modules}

    \bigskip

    \begin{tikzpicture}[scale=.8]
      \draw (0,0) node[rond7] (z1)  {$\zero j$};
      \draw[latex'-] (z1.-135) -- +(-135:3mm);
      \draw (2,0) node[rond] (a1) {};
      \draw (4,0) node[rond] (b1) {};
      \draw (6,0) node[lrond7] (o1) {$\one j$};
      \draw (8,0) node[rond] (c1) {};
      \draw (10,0) node[rond] (d1) {};
      \draw (12,0) node[lrond7] (z'1) {$\zero j'$};
      \path (z1) edge[-latex',bend right] node[below] {\receive{i-ask$_j$}} (a1);
      \path (a1) edge[-latex',bend right] node[above] {\receive{i-ok$_j$}} (z1);
      \path (a1) edge[-latex'] node[above] {\broadcast{i-ack$_j$}} (b1);
      \path (b1) edge[-latex'] node[above] {\receive{i-ok$_j$}} (o1);
      \path (o1) edge[-latex',bend right] node[below] {\receive{d-ask$_j$}} (c1);
      \path (c1) edge[-latex',bend right] node[above] {\receive{d-ok$_j$}} (o1);
      \path (c1) edge[-latex'] node[above] {\broadcast{d-ack$_j$}} (d1);
      \path (d1) edge[-latex'] node[above] {\receive{d-ok$_j$}} (z'1);
      \path (o1) edge[out=-90, in=-150,looseness=6,-latex'] node[below]
            {\receive{i-init}, \receive{i-ask$_*$}, \receive{i-ok$_*$},
              \receive{i-exit}, \receive{d-exit}} (o1); 
    \end{tikzpicture}
    \caption{The part of the protocol for counter nodes}
    \label{fig-counter}

    \bigskip

    \begin{tikzpicture}
      \begin{scope}
      \draw (0,0) node[rond7] (aux1) {\free 1};
      \draw[latex'-] (aux1.-135) -- +(-135:2.4mm);
      \draw (2,0) node[rond] (a1) {};
      \draw (4,0) node[lrond7] (endaux1) {\done 1};
      \path (aux1) edge[-latex'] node[above] {\receive{i-init}} (a1);
      \path (a1) edge[-latex'] node[above] {\broadcast{\maux$_1$}} (endaux1);
      \end{scope}
      \begin{scope}[xshift=-1cm,yshift=-1cm]
        \draw (0,0) node[rond7] (aux2) {\free 2};
        \draw[latex'-] (aux2.-135) -- +(-135:2.4mm);
        \draw (2,0) node[rond] (b1) {};
        \draw (4,0) node[rond] (b2) {};
        \draw (6,0) node[lrond7] (endaux2) {\done 2};
        \path (aux2) edge[-latex'] node[above] {\receive{i-ask$_j$}} (b1);
        \path (b1) edge[-latex'] node[above] {\receive{i-ok$_j$}} (b2);
        \path (b2) edge[-latex'] node[above] {\broadcast{\maux$_2$}} (endaux2);
      \end{scope}
      \begin{scope}[yshift=-2cm]
        \draw (0,0) node[rond7] (aux3)  {\free 3};
        \draw[latex'-] (aux3.-135) -- +(-135:2.4mm);
        \draw (2,0) node[rond] (c1) {};
        \draw (4,0) node[lrond7] (endaux3)  {\done 3};
        \path (aux3) edge[-latex'] node[above] {\receive{i-ok$_j$}} (c1);
        \path (c1) edge[-latex'] node[above] {\broadcast{\maux$_3$}} (endaux3);
      \end{scope}
      \begin{scope}[yshift=-3cm]
        \draw (0,0) node[rond7] (aux5)  {\free 5};
        \draw[latex'-] (aux5.-135) -- +(-135:2.4mm);
        \draw (2,0) node[rond] (e1) {};
        \draw (4,0) node[lrond7] (endaux5)  {\done 5};
        \path (aux5) edge[-latex'] node[above] {\receive{d-ok$_j$}}
          node[below] {\receive{t-ok$_j$}} (e1);
        \path (e1) edge[-latex'] node[above] {\broadcast{\maux$_5$}} (endaux5);
      \end{scope}
      \begin{scope}[xshift=8cm,xscale=1.7]
        \draw (0,0) node[rond7] (aux4)  {\free 4};
        \draw[latex'-] (aux4.135) -- +(135:1.8mm);
        \draw (-1,-1) node[rond] (z1) {};
        \draw (0,-1) node[rond] (y1) {};
        \draw (1,-1) node[rond] (x1) {};
        \draw (-1,-2) node[rond] (z2) {};
        \draw (0,-2) node[rond] (y2) {};
        \draw (1,-2) node[rond] (x2) {};
        \draw (0,-3) node[lrond7] (endaux4) {\done 4};
        \path (aux4) edge[-latex'] node[left] {\receive{t-ask$_1$}} (z1);
        \path (aux4) edge[-latex'] node[right] {\receive{t-ask$_2$}} (x1);
        \path (aux4) edge[-latex'] node {\receive{d-ask$_j$}} (y1);
        \path (z1) edge[-latex'] node[] {\broadcast{t-ack$_1$}} (z2);
        \path (x1) edge[-latex'] node[] {\broadcast{t-ack$_2$}} (x2);
        \path (y1) edge[-latex'] node {\receive{d-ok$_j$}} (y2);
        \path (z2) edge[-latex'] node[left] {\receive{t-ok$_1$}} (endaux4);
        \path (x2) edge[-latex'] node[right] {\receive{t-ok$_2$}} (endaux4);
        \path (y2) edge[-latex'] node {\broadcast{\maux$_4$}} (endaux4);
      \end{scope}
    \end{tikzpicture}
    \caption{Parts of the protocol for auxiliary nodes}
    \label{fig-aux}
    
  \end{figure}


  Let us explain the intended behaviour of the incrementation module
  of Fig.~\ref{fig-modules}: when entering the module, our control
  node~$\node$ in state~$L$ is linked to $c_1$ counter nodes in
  state~$\one 1$ and to~$c_2$ counter nodes in state~$\one 2$; it~has
  no other links. Moreover, all auxiliary nodes are either in
  state~$\free i$ or in state~$\done i$. Running through the
  incrementation module from~$L$ will use one counter node~$\nodem$ in
  state~$\zero j$ (which is used to effectively encode the increase of
  counter~$c_j$) and four auxiliary nodes~$a_1$ (initially in
  state~$\free 1$), $a_2$ (in~state~$\free 2$), and $a_3$ and~$a'_3$
  (in~state~$\free 3$).

  The execution then runs as follows:
  \begin{itemize}
  \item a link is created between the control node~$\node$ and the first
    auxiliary node~$a_1$, followed by a message
    exchange~\broadcast{i-init};
  \item a link is created between~$\node$ and~$\nodem$, and node~$a_1$
    broadcasts~\broadcast{\maux$_1$};
  \item a link is created between~$\node$ and~$a_2$, and $\node$ broadcasts
    \broadcast{i-ask$_j$}, which is received by both~$a_2$ and~$\nodem$;
  \item a link is created between~$\node$ and~$a_3$; node~$\nodem$ sends its
    acknowledgement~\broadcast{i-ack$_j$} to~$\node$;
  \item a link is created between~$\node$ and~$a'_3$; node~$\node$ sends
    \broadcast{i-ok$_j$}, received by~$\nodem$, $a_2$, $a_3$ and~$a'_3$;
  \item the link between~$\node$ and~$a_1$ is removed, and $a_2$
    sends~\broadcast{\maux$_2$};
  \item the link between~$\node$ and~$a_2$ is removed, and $a_3$
    sends~\broadcast{\maux$_3$};
  \item the link between~$\node$ and~$a_3$ is removed, and $a'_3$
    sends~\broadcast{\maux$_3$};
  \item finally, the link between~$\node$ and~$a'_3$ is removed, and $\node$
    sends~\broadcast{i-exit}.
  \end{itemize}
  Figure~\ref{fig-steps} depicts this sequence of steps graphically.
  
\Conefalse
\Ctwofalse
\Cthreefalse
\Cfourfalse
\Cfivefalse

\def\step#1#2#3#4#5#6#7{%
  \draw[dotted,rounded corners=3mm] (0,1.7) -| (2,-1.7) node[coordinate,pos=.75] (front) {} -| (-2,1.7) node[midway,coordinate] (rear) {} -- (0,1.7);
  \if\relax#7\relax
    \path (front) edge[-latex',dotted] +(1,0);
  \fi
  \draw (0,0) node[rond] (L) {$#1$};
  \draw (-45:1.5cm) node[oval] (C) {$#2$};
  \draw (45:1.5cm) node[oval] (C') {$\one j$};
  \draw (15:1.5cm) node[oval] (C'') {$\one j$};
  \path (L) edge[dashed] (C');\path (L) edge[dashed] (C'');
  \draw (-120:1.5cm) node[oval] (a1) {$#3$};
  \draw (-160:1.5cm) node[oval] (a2) {$#4$};
  \draw (160:1.5cm) node[oval] (a3) {$#5$};
  \draw (120:1.5cm) node[oval] (a4) {$#6$};
  \ifCone\draw[dashed] (L) -- (C);\fi
  \ifCtwo\draw[dashed] (L) -- (a1);\fi
  \ifCthree\draw[dashed] (L) -- (a2);\fi
  \ifCfour\draw[dashed] (L) -- (a3);\fi
  \ifCfive\draw[dashed] (L) -- (a4);\fi
}

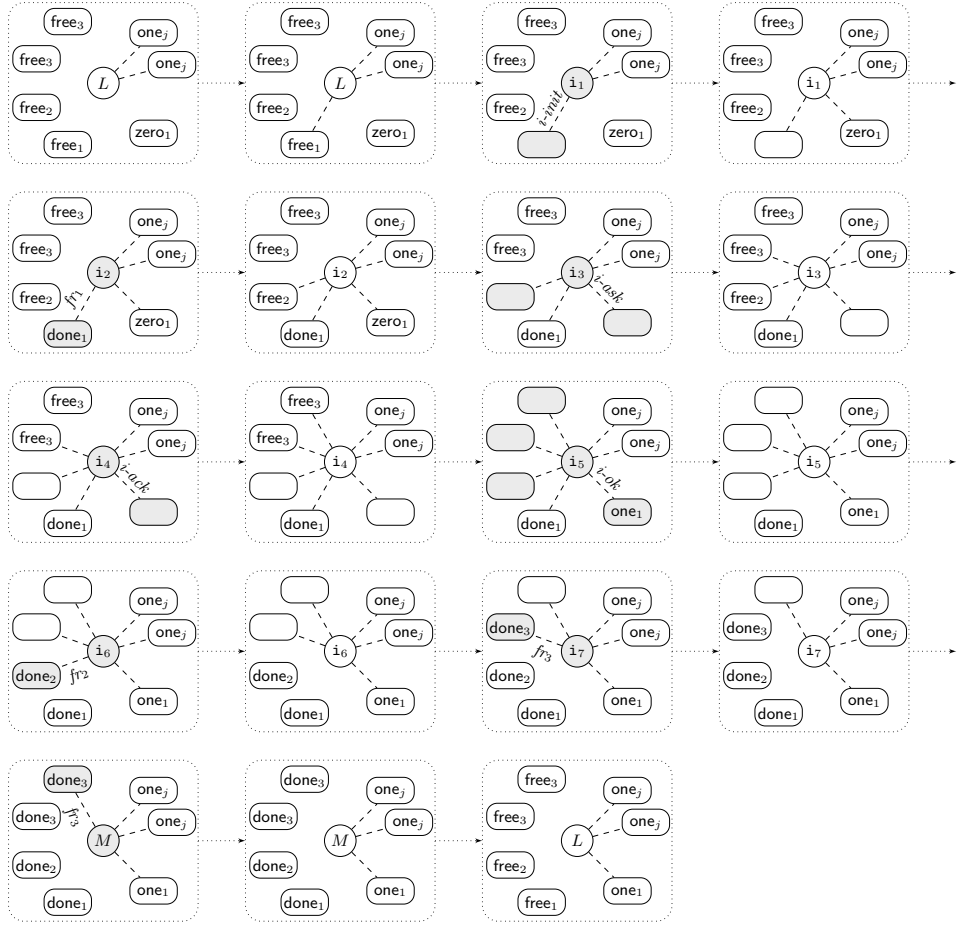
\begin{figure}[htp]
  \scalebox{.7}{%
  \begin{tikzpicture}[scale=.9]
    \path[use as bounding box] (-2,0) -- +(16.5,-18);
    \tikzstyle{oval}=[draw,rounded corners=2mm,inner sep=1pt,minimum width=9mm,minimum height=5mm]
    \begin{scope}
    \step L{\zero 1}{\free 1}{\free 2}{\free 3}{\free 3}{}
    \end{scope}
    \Ctwotrue
    \begin{scope}[xshift=5cm]
    \step {L}{\zero 1}{\free 1}{\free 2}{\free 3}{\free 3}{}
    \end{scope}
    \begin{scope}[xshift=10cm]
    \step {\inc1}{\zero 1}{}{\free 2}{\free 3}{\free 3}{}
    \path (L) -- (a1) node[midway,above,sloped] {\textit{i-init}};
    \draw (L) node[rond,opacity=.1,fill=black!80!white] {};
    \draw (a1) node[oval,opacity=.1,fill=black!80!white] {};
    \end{scope}
    \Conetrue
    \begin{scope}[xshift=15cm]
    \step {\inc1}{\zero 1}{}{\free 2}{\free 3}{\free 3}{}
    \end{scope}
    \begin{scope}[yshift=-4cm]
      \begin{scope}
        \step {\inc2}{\zero 1}{\done 1}{\free 2}{\free 3}{\free 3}{}
        \path (L) -- (a1) node[midway,above,sloped] {\textit{\maux$_1$}};
        \draw (L) node[rond,opacity=.1,fill=black!80!white] {};
        \draw (a1) node[oval,opacity=.1,fill=black!80!white] {};
      \end{scope}
      \global\Cthreetrue
      \begin{scope}[xshift=5cm]
        \step {\inc2}{\zero 1}{\done 1}{\free 2}{\free 3}{\free 3}{}
      \end{scope}
      \begin{scope}[xshift=10cm]
        \step {\inc3}{}{\done1}{}{\free 3}{\free 3}{}
        \path (L) -- (C) node[midway,above,sloped] {\textit{i-ask}};
        \draw (L) node[rond,opacity=.1,fill=black!80!white] {};
        \draw (C) node[oval,opacity=.1,fill=black!80!white] {};
        \draw (a2) node[oval,opacity=.1,fill=black!80!white] {};
      \end{scope}
      \global\Cfourtrue
      \begin{scope}[xshift=15cm]
        \step {\inc3}{}{\done1}{\free 2}{\free 3}{\free 3}{}
      \end{scope}
    \end{scope}
    \begin{scope}[yshift=-8cm]
      \begin{scope}
        \step {\inc4}{}{\done1}{}{\free 3}{\free 3}{}
        \path (L) -- (C) node[midway,above,sloped] {\textit{i-ack}};
        \draw (L) node[rond,opacity=.1,fill=black!80!white] {};
        \draw (C) node[oval,opacity=.1,fill=black!80!white] {};
      \end{scope}
      \global\Cfivetrue
      \begin{scope}[xshift=5cm]
        \step {\inc4}{}{\done 1}{}{\free 3}{\free 3}{}
      \end{scope}
      \begin{scope}[xshift=10cm]
        \step {\inc5}{\one1}{\done1}{}{}{}{}
        \path (L) -- (C) node[midway,above,sloped] {\textit{i-ok}};
        \draw (L) node[rond,opacity=.1,fill=black!80!white] {};
        \draw (C) node[oval,opacity=.1,fill=black!80!white] {};
        \draw (a2) node[oval,opacity=.1,fill=black!80!white] {};
        \draw (a3) node[oval,opacity=.1,fill=black!80!white] {};
        \draw (a4) node[oval,opacity=.1,fill=black!80!white] {};
      \end{scope}
      \global\Ctwofalse
      \begin{scope}[xshift=15cm]
        \step {\inc5}{\one 1}{\done1}{}{}{}{}
      \end{scope}
    \end{scope}
    \begin{scope}[yshift=-12cm]
      \begin{scope}
        \step {\inc6}{\one1}{\done1}{\done2}{}{}{}
        \path (L) -- (a2) node[midway,below,sloped] {\textit{\maux$_2$}};
        \draw (L) node[rond,opacity=.1,fill=black!80!white] {};
        \draw (a2) node[oval,opacity=.1,fill=black!80!white] {};
      \end{scope}
      \global\Cthreefalse
      \begin{scope}[xshift=5cm]
        \step {\inc6}{\one1}{\done 1}{\done2}{}{}{}
      \end{scope}
      \begin{scope}[xshift=10cm]
        \step {\inc7}{\one1}{\done1}{\done2}{\done3}{}{}
        \path (L) -- (a3) node[midway,below,sloped] {\textit{\maux$_3$}};
        \draw (L) node[rond,opacity=.1,fill=black!80!white] {};
        \draw (a3) node[oval,opacity=.1,fill=black!80!white] {};
      \end{scope}
      \global\Cfourfalse
      \begin{scope}[xshift=15cm]
        \step {\inc7}{\one 1}{\done1}{\done2}{\done 3}{}{}
      \end{scope}
    \end{scope}
    \begin{scope}[yshift=-16cm]
      \begin{scope}
        \step {M}{\one1}{\done1}{\done2}{\done3}{\done3}{}
        \path (L) -- (a4) node[midway,below,sloped] {\textit{\maux$_3$}};
        \draw (L) node[rond,opacity=.1,fill=black!80!white] {};
        \draw (a4) node[oval,opacity=.1,fill=black!80!white] {};
      \end{scope}
      \global\Cfivefalse
      \begin{scope}[xshift=5cm]
        \step {M}{\one1}{\done 1}{\done2}{\done3}{\done3}{}
      \end{scope}
      \begin{scope}[xshift=10cm]
        \step L{\one 1}{\free 1}{\free 2}{\free 3}{\free 3}{n}
      \end{scope}
    \end{scope}
  \end{tikzpicture}}
  \caption{The 18 steps of an incrementation of counter~$c_1$}
  \label{fig-steps}
\end{figure}
 
After this sequence of steps, node~$\node$ has an
  extra link to a counter node in state~$\one j$, which indeed
  corresponds to incrementing counter~$c_j$. Moreover, no nodes have
  been left in an intermediary state.
  A~similar analysis can be done for the second module, which
  implements the zero-test and decrementation. This way, we~can prove
  that if the two-counter machine has a halting computation, then
  there is an initial configuration of our broadcast protocol from
  which there is an execution synchronizing in the set $F$ formed of
  the halting control state and states $\one j$, $\zero j'$ and~$\done
  i$.

  \medskip It~now remains to prove the other direction. More
  precisely, we prove that from a $1$-constrained synchronizing
  execution of the protocol, we~can extract a synchronizing execution
  in some normal form, from which we derive a halting execution of the
  two-counter machine.

  \looseness=-1 Fix a $1$-constrained synchronizing execution of the
  broadcast network.  First notice that when a control node~$\node$
  reaches some state~$L$ (the~first node of an incrementation or
  decrementation module), it~may only be linked to counter nodes in
  state~$\one j$: this~is because states~$L$ can only be reached by
  sending~\broadcast{i-exit}, \broadcast{d-exit},
  \broadcast{t-exit}, or \broadcast{start}. The~former two cases
  may only synchronize with counter nodes in state~$\one j$;
  in~the~other two cases, node~$\node$ may be linked to no other node.
  Hence, for a
  control node~$\node$ to traverse an incrementation module, it~must
  get links to four auxiliary nodes (in order to receive the four
  $\maux$ messages), those four links must be removed (to~avoid
  reaching the sink state), and an extra link has to be created in
  order to receive message \textit{i-ack$_j$}. In~total, traversing an
  incrementation module takes nine communication steps and at least
  nine reconfiguration steps. Similarly, traversing a decrementation
  module via any of the two branches takes at least as many
  reconfiguration steps as communication steps.  In~the end, taking
  into account the initial~$\broadcast{start}$ communication step, if
  a control node~$\node$ is involved in $B_\node$ communication steps,
  it~must be involved in at least $B_\node-1$ reconfiguration steps.

  Assume that every control node~$\node$ is involved in at least
  $B_\node$ reconfiguration steps: then we would have at least as many
  reconfiguration steps as communication steps, which in a
  $1$-constrained execution is impossible. Hence there must be a control
  node~$\node_0$ performing $B_{\node_0}$ communication steps and
  exactly $B_{\node_0}-1$ reconfiguration steps. As~a consequence,
  when traversing an incrementation module, node~$\node_0$ indeed gets
  connected to exactly one new counter node, which indeed must be in
  state~$\one j$ when $\node_0$ reaches the first state of the next
  module. Similarly, traversing a decrementation\slash zero-test
  module indeed performs the expected changes. It~follows that the
  sequence of steps involving node~$\node_0$ encodes a halting
  execution of the two-counter machines.
  \qed

  \medskip
  The $1$-constrained executions in the proof of
  Theorem~\ref{thm-undec} have the additional property that all graphs
  describing configurations are $2$-bounded-path configurations. For
  $K \in \nats$ a configuration $\graph$ is a \emph{$K$-bounded-path
    configuration} if the length of all simple paths in $\graph$ is
  bounded by~$K$. Note that a constant bound on the length of simple
  paths implies that the diameter (\emph{i.e.}~the length of the
  longest shortest path between any pair of vertices) is itself
  bounded. The~synchronization problem was proved to be undecidable
  for broadcast networks \emph{without reconfiguration} when
  restricting to $K$-bounded-path
  configurations~\cite{DSZ10}. In~comparison, for reconfigurable
  broadcast networks under $k$-constrained reconfigurations, the
  undecidability result stated in Theorem~\ref{thm-undec} can be
  strengthened into:
\begin{corollary}
\label{coro-undec-bounded}
The synchronization problem is undecidable for reconfigurable
broadcast networks under $k$-constrained reconfigurations when
restricted either to bounded-path configurations, or to
bounded-diameter configurations.
\end{corollary}
\end{proof}

\subsection{Decidability results}

\paragraph{$f$-constrained and $k$-locally-constrained reconfigurations.}
From the equivalence (w.r.t. synchronization) of
$k$-locally-constrained, $f$-constrained and unconstrained executions
(Lemmas~\ref{lemma-1loc} and~\ref{lemma-fconst}), and thanks to
Theorem~\ref{thm-mobile}, we~immediately get:
\begin{corollary}
  Let $k \in \nats$ and $f\colon \nats \to \nats$ be a non-decreasing
  diverging function.  The synchronization problem for reconfigurable
  broadcast networks under $k$-locally-\hskip0pt constrained
  (resp.~$f$-constrained) reconfigurations is decidable in \PTIME.
\end{corollary}

\paragraph{Bounded degree topology.}
We now return to $k$-constrained reconfigurations, and explore restrictions 
that allow one to recover decidability of the synchronization problem.
We~further restrict $k$-constrained reconfigurations by requiring that
the degree of nodes remains bounded, by $1$; in other terms,
communications correspond to \emph{rendez-vous} between the
broadcasting node and its single neighbour.

\begin{restatable}{theorem}{thmoneb}
  \label{thm-1b}
  The synchronization problem is decidable for reconfigurable
  broadcast networks under
  $k$-constrained reconfiguration when restricted to 1-bounded-degree
  topologies.
\end{restatable}

  \begin{proof}
  The proof is by reduction to the reachability problem for Petri
  nets, which is known to be decidable~\cite{May81,Kos82,Ler12}. Let
  $\BP = (\States,\state_0,\Mess,\Trans)$ be a broadcast protocol,
  where $\States = \{q_1,q_2,\cdots,q_n\}$.  With~$\BP$, we~associate
  a Petri net~$\calN$ running in three phases: the
  \emph{initialization} phase sets the net into a marking encoding
  some initial configuration of the network; during the
  \emph{simulation} phase, $\calN$~simulates the execution of the
  reconfigurable broadcast network; finally, in a \emph{checking}
  phase, it~will make sure that all nodes have reached a target state.

  The Petri net $\calN$ has three types of nodes:
  \begin{itemize}
  \item for each state~$q_i\in\States$, a place~$p_i$ is used to
    represent \emph{isolated} nodes (\emph{i.e.}\ having no neighbour
    in the communication graph) in state~$q_i$. In~the simulation
    phase of our encoding, the~number of tokens in~$p_i$ will
    correspond to the number of isolated nodes in state~$q_i$;
  \item for any pair of states~$(q_i,q_j)$ (possibly with~$q_i=q_j$),
    a place~$p_{i,j}$ is used to represent pairs of connected nodes,
    one in state~$q_i$ and the other in state~$q_j$. Since we consider
    1-bounded-degree topologies, those nodes may not have additional
    connections. Again, the simulation phase of our encoding will
    maintain the correspondence between the number of links between
    nodes in~$q_i$ and~$q_j$ and the number of tokens in
    place~$p_{i,j}$;
  \item finally, $\calN$ has $k+4$ auxiliary places: four places
    $\pstart$, $\psimul$, $\pcheck$ and~$\pend$ indicate the different
    phases, while the other $k$ places
    $\{\preconf i \mid 1\leq i\leq k\}$ are used to constrain the
    number of reconfigurations.
  \end{itemize}

  The transitions can be split in three groups, according to the
  phases they correspond to:
  \begin{itemize}
  \item in~order to initiate the simulation, transitions from
    place~$\pstart$ will place tokens in places~$p_0$ and~$p_{0,0}$,
    thereby encoding an initial configuration of the network. This is
    performed via three transitions:
    \begin{itemize}
    \item transition $t^\phinit_1$ with $\LB{t^\phinit_1}=\{\pstart\}$ and
      $\RB{t^\phinit_1} = \{\pstart,p_0\}$;
    \item transition $t^\phinit_2$ with $\LB{t^\phinit_2}=\{\pstart\}$ and
      $\RB{t^\phinit_2} = \{\pstart,p_{0,0}\}$;
    \item transition $t^\phinit_{\phsimul}$ with
      $\LB{t^\phinit_\phsimul}=\{\pstart\}$ and
      $\RB{t^\phinit_\phsimul}=\{\psimul\}$.
    \end{itemize} 

  \item the simulation phase starts when there is a token
    in~$\psimul$; it~then proceeds by simulating communication steps
    and reconfiguration steps, alternately.

    Communication steps are easily simulated:
    \begin{itemize}
    \item either an isolated node in some state~$q_i$ performs a
      transition $\delta=(q_i,\broadcast a, q_j)$, which can be
      simulated by a transition $t^{\phsimul}_{\delta}$ with
      $\LB{t^{\phsimul}_{\delta}}=\{\psimul,p_i\}$ and
      $\RB{t^{\phsimul}_{\delta}}=\{\preconf1,p_j\}$. Place~$\preconf1$
      indicates that the next step in the simulation will be the first
      (of possibly up to~$k$) reconfiguration step;
    \item a communication step may also take place between two
      connected nodes. Then for any two transitions
      $\delta=(q_i,\broadcast a, q_j)$ and $\delta'=(q_m,\receive a,
      q_n)$, we have a transition $t^{\phsimul}_{\delta,\delta'}$ with
      $\LB{t^{\phsimul}_{\delta,\delta'}} = \{\psimul, p_{i,m}\}$ and
      $\RB{t^{\phsimul}_{\delta,\delta'}} = \{\preconf1, p_{j,n}\}$.
    \end{itemize}
    When a token is in a reconfiguration place, reconfiguration steps
    may take place. Atomic reconfigurations may either create links
    between two isolated nodes, or break links between connected
    nodes:
    \begin{itemize}
    \item for any pair of states~$q_i$ and~$q_j$, we~have transitions
      $t^{\phreconf_m}_{i\leftrightarrow j}$, for $1\leq m\leq k$,
      creating a new link between nodes in state~$q_i$ and~$q_j$. Such
      transitions may occur at any of the $k$ atomic steps of
      reconfigurations:
      $\LB{t^{\phreconf_m}_{i\leftrightarrow j}} = \{\preconf m, p_i,
      p_j\}$ and
      $\RB{t^{\phreconf_m}_{i\leftrightarrow j}} = \{\preconf{m+1},
      p_{i,j}\}$, for any $1\leq m\leq k$ (with the convention that
      $\preconf{k+1}= \psimul$);
    \item conversely, breaking a link between nodes in states~$q_i$
      and~$q_j$ is encoded by transitions
      $t^{\phreconf_m}_{i\not\leftrightarrow j}$ such that
      $\LB{t^{\phreconf_m}_{i\not\leftrightarrow j}} = \{\preconf m, p_{i,j}\}$
      and $\RB{t^{\phreconf_m}_{i\not\leftrightarrow j}} =
      \{\preconf{m+1}, p_{i}, p_{j}\}$, for any $1\leq m\leq k$.
    \item finally, we~allow reconfigurations of less than $k$ links by
      adding transitions from $\preconf m$ to~$\psimul$, for each~$m$:
      $\LB{t^{\phreconf_m}_{\textit{end}}}=\{\preconf m\}$ and
      $\RB{t^{\phreconf_m}_{\textit{end}}}=\{\psimul\}$.

      We~also allow ending the simulation phase by adding a transition
      that moves the token from~$\psimul$ to~$\pcheck$.
    \end{itemize}

  \item In the checking phase, we aim at checking that all nodes are
    in a target state. To this aim, place~$\pcheck$ absorbs all tokens
    in places that correspond to target states, that is places~$p_i$
    whose associated state~$q_i$ is a target state, as well as
    places~$p_{i,j}$ for which both $q_i$ and~$q_j$ are target
    states. Hence, for each~$q_i\in\targetset$, we~have a transition
    $t^{\phcheck}_{i}$ with $\LB{t^{\phcheck}_i} = \{\pcheck, p_i\}$
    and $\RB{t^{\phcheck}_i} = \{\pcheck\}$. Similarly, for any two
    states~$q_i$ and~$q_j$ in~$\targetset$, we have a transition
    $t^{\phcheck}_{i,j}$ with
    $\LB{t^{\phcheck}_{i.j}} = \{\pcheck, p_{i,j}\}$ and
    $\RB{t^{\phcheck}_{i,j}} = \{\pcheck\}$.
  \end{itemize}

  By construction, the reconfigurable broadcast network defined by the
  broadcast protocol~$\BP$ has a $k$-constrained $1$-bounded-degree
  execution from some initial configuration if, and only~if, the Petri
  net~$\calN$ has an execution from the initial marking with a single
  token in $\pstart$ to the marking with a single token in
  $\pend$. 
We thus reduced to reachability of Petri
  nets, hence yielding decidability.\qed
  \end{proof}

\section{Conclusion}
Restricting reconfigurations in reconfigurable broadcast networks is
natural to better reflect mobility when communications are frequent
enough and the movement of nodes is not chaotic. In this paper, we
studied how constraints on the number of reconfigurations (at each
step and for each node, at each step and globally, or along an
execution) change the semantics of networks, in particular with
respect to the synchronization problem, and affect its
decidability. Our~main results are the equivalence of $k$-constrained
and $k$-balanced semantics, the~undecidability of synchronization
under $k$-constrained reconfigurations, and its decidability when
restricting to $1$-bounded-degree topologies.

\enlargethispage{5mm}
As future work, we propose to investigate, beyond the coverability and
synchronization problems, richer objectives such as cardinality
reachability problems as in~\cite{DSTZ12}. Moreover, for semantics
with constrained reconfigurations that are equivalent to the
unconstrained one as far as the coverability and synchronization
problems are concerned, it would be worth studying the impact of the
reconfiguration restrictions (\emph{e.g.}\ $k$-locally-constrained or
$f$-constrained) on the minimum number of nodes for which a
synchronizing execution exists, and on the minimum number of steps to
synchronize.



\end{document}